\documentclass[review,11pt]{elsarticle}

\usepackage[english]{babel}

\usepackage[a4paper,top=2cm,bottom=2cm,left=3cm,right=3cm,marginparwidth=1.75cm]{geometry}
\usepackage{amsmath}
\usepackage{graphicx}
\usepackage[colorlinks=true, allcolors=blue]{hyperref}
\usepackage{comment}
\usepackage{float}

\usepackage{amssymb,amsthm,booktabs,caption,enumerate,eurosym,graphicx,hyperref,mathrsfs,mathtools,times,xcolor, subcaption}

\newcommand{\todo}[1]{\textcolor{red}{#1}}
\newcommand{\xta}{x_{\textrm{TA}}}
\newcommand{\xp}{x_{\textrm{P}}}
\newcommand{\xtn}{x_{\textrm{TN}}}
\newcommand{\xtt}{x_{\textrm{TT}}}
\newcommand{\xws}{x_{\textrm{WS}}}
\newcommand{\xd}{x_{\textrm{D}}}
\newcommand{\Pmax}{p_{\textrm{MAX}}}

\newcommand{\xsta}{x^*_{\textrm{TA}}}
\newcommand{\xsp}{x^*_{\textrm{P}}}
\newcommand{\xstn}{x^*_{\textrm{TN}}}
\newcommand{\xstt}{x^*_{\textrm{TT}}}
\newcommand{\xsws}{x^*_{\textrm{WS}}}
\newcommand{\xsd}{x^*_{\textrm{D}}}

\newcommand{\xrp}{\tilde{x}_{\textrm{P}}}
\newcommand{\xr}{\tilde{x}}

\newcommand{\avg}{\overline{x}}
\newcommand{\lb}{\mathbf{l}}
\newcommand{\ub}{\mathbf{u}}

\newcommand{\maxtchange}{m_{\delta T}}

\begin{document}
\begin{frontmatter}

\title{Counterfactual optimization for fault prevention in complex wind energy systems}

\author[EC]{Emilio Carrizosa}
\ead{ecarrizosa@us.es}
\author[MF]{Martina Fischetti}
\ead{mfischetti@us.es}
\author[RH]{Roshell Haaker}
\ead{roshell.haaker@vattenfall.com}
\author[JMM]{Juan Miguel Morales}
\ead{juan.morales@uma.es}
\address[EC]{Department of Statistics and Operations Research, University of Seville, Spain}
\address[MF]{Department of Statistics and Operations Research, University of Seville, Spain}
\address[RH]{Vattenfall BA Wind, the Netherlands}
\address[JMM]{Department of Statistics, Operations Research and Applied Mathematics, University of Malaga, Spain}

\begin{abstract}
Machine Learning models are increasingly used in businesses to detect faults and anomalies in complex systems. In this work, we take this approach a step further: beyond merely detecting anomalies, we aim to identify the optimal control strategy that restores the system to a safe state with minimal disruption. We frame this challenge as a counterfactual problem: given a Machine Learning model that classifies system states as either ``good" or ``anomalous," our goal is to determine the minimal adjustment to the system’s control variables (i.e., its current status) that is necessary to return it to the ``good" state.

To achieve this, we leverage a mathematical model that finds the optimal counterfactual solution while respecting system-specific constraints. Notably, most counterfactual analysis in the literature focuses on individual cases where a person seeks to alter their status relative to a decision made by a classifier—such as for loan approval or medical diagnosis. Our work addresses a fundamentally different challenge: optimizing counterfactuals for a complex energy system, specifically an offshore wind turbine oil-type transformer. This application not only advances counterfactual optimization in a new domain but also opens avenues for broader research in this area. 

Our tests on real-world data provided by our industrial partner show that our methodology easily adapts to user preferences and brings savings in the order of 3 million \euro~ per year in a typical farm.
\end{abstract}

\end{frontmatter}

\section{Introduction} 
Energy systems are becoming increasingly more complex, making it more challenging---and more critical---to detect faults early and develop strategies to mitigate them. Offshore wind turbines, in particular, present a significant concern, as faults can lead to substantial production losses (since turbines are typically shut down for repairs when a fault is detected) and safety risks (due to the difficulty of reaching turbines located far offshore, especially in adverse weather conditions). In this context, Machine Learning (ML) techniques have become an industry standard for early fault detection \citep{MLFaults}. 

Energy companies can monitor various sensor readings from the turbines and apply ML methods to identify potential issues with components. In this paper, we define a \textit{fault} (or \emph{faulty state}) as a condition where a component is in an unsafe status, while an \textit{anomaly} refers to any irregularity that is not necessarily dangerous. Note that faults are a subset of anomalies. When a fault is detected, a controller is immediately activated to prevent severe damage to the turbine. Machine Learning models can detect anomalies in advance, providing companies with a window of time to intervene before faults occur. 

Nevertheless, as we will discuss in more detail, it is far from straightforward to determine how to address \textit{anomalies}. In this paper, we use Mathematical Optimization to inform these decisions. Specifically, we introduce a novel approach that combines Operations Research and counterfactual analysis to suggest an optimized control strategy for restoring a component to a healthy state. To the best of our knowledge, this is the first application of these techniques for fault prevention.

The specific context of fault prevention for oil-type transformers in offshore wind turbines drives us to explore new areas in counterfactual optimization with broader theoretical significance. For example, as we will see later in the paper, we incorporate nested machine learning models within the constraints of a mixed-integer quadratic programming model (Section~\ref{subsec:controller simulation}), account for cost in our counterfactual objective function (Section~\ref{sec:cost-opt}), and examine the interesting relationship between ML model accuracy and counterfactual control strategies (Section~\ref{sec:accuracy&CE}). These methods are broadly applicable to a wide range of complex systems.

\subsection{A short introduction to Counterfactual Analysis}

Our work offers a fresh perspective on optimization within the context of Counterfactual Analysis. Once a classification model is trained and an instance is classified as ``bad", Counterfactual Analysis is employed to determine how records need to be modified in their features to be classified under the desired ``good" category. A classic illustration is loan approval, where an individual submits a list of personal details to the bank, which would be referred to as \emph{features} in Machine Learning, and the bank employs a Machine Learning algorithm to categorize them as either ``good" or ``bad" payers for a loan. When someone is classified as ``bad" and their loan application is rejected, they may want to consider how to alter their circumstances to be reclassified as a ``good" credit risk. This process is known as Counterfactual Analysis (for further details, the reader may refer to \citep{artelt2019},\citep{Guidotti24}, \citep{KarimiBBV20} for comprehensive surveys on the topic).

 As shown in \citep{CARRIZOSA2024}, identifying counterfactual explanations can be framed as solving a mathematical optimization problem. The objective is to minimize the number of changes to an instance’s features while maximizing the probability of it being classified as “good,” all while adhering to various constraints. The literature offers a variety of numerical approaches to efficiently solve this optimization problem. These include smooth optimization methods, such as \citep{joshi2019} and \citep{Ramakrishnan}; mixed-integer optimization techniques, as seen in \citep{CARRIZOSA2024}, \citep{FischettiJo}, \citep{kanamori_2020}, and \citep{Kanamori_2021}; multi-objective optimization strategies, as discussed in \citep{Dandl2020}, \citep{delser}, and \citep{Raimundo}; robust optimization approaches, like those in \citep{Marango_Kurtz}; SAT-based methods, e.g., \citep{kanamori_2020}; and heuristic and metaheuristic techniques, as explored in \citep{Guidotti2019}.

In this work, we extend the application of counterfactual analysis to a new domain: fault prevention in complex energy systems. Similar to how a loan applicant provides personal data for evaluation, a wind farm component records various measurements (its features). It is becoming industry standard to use Machine Learning classifiers to analyze these features and assess whether the component is in a healthy state. Our objective is to go a step further and, once the status of the component is classified as ``anomalous", identify the optimal counterfactual control strategy, i.e., one that requires minimal adjustments to the component’s status while ensuring it is reclassified as ``healthy" by the Machine Learning classifier.

\subsection{Fault detection and prevention in complex energy systems}

Energy generation and distribution systems are inherently complex, with multiple interconnected components interacting in intricate ways. Within this framework, detecting and preventing faults pose significant engineering challenges. 
Consider, for example, an offshore wind farm: a highly complex energy system in which numerous components interact through complex dynamics (such as wind speed, waves, turbulence, aerodynamics, and temperature), and through sophisticated engineering subsystems (including electrical circuits, aerodynamic structures, and control mechanisms). Not only is this system exceptionally complex, but it is also extremely costly.
As demonstrated in \cite{FISCHETTI2021EjorSurvey} and \cite{FISCHETTIEdelman}, the use of optimization in offshore wind energy can have a significant economic impact, reducing costs and making wind energy more competitive. This, in turn, contributes to a more sustainable and environmentally friendly future.
Given the complexity of such systems, preventing faults becomes highly challenging and costly. The cost (and safety hazard) is particularly high for offshore wind turbines. Offshore personnel deployment requires, indeed, specialized vessels or helicopters, depending on weather conditions, and is associated with significant safety risks.

Furthermore, in most cases, the turbine's production must be reduced (often to zero) until an inspection is conducted and maintenance is completed. This process can take days, especially in poor weather conditions, resulting in significant financial losses for the company, whose primary business is energy sales.

As a result, two key factors emerge: (i) it is crucial for energy companies to reliably identify faults and minimize false alarms, and (ii) acting swiftly with a well-defined strategy to address or mitigate faults and anomalies can have a profound impact. For (i), many companies now use Machine Learning (ML) to detect anomalies more accurately \citep{faultDiag, MLFaults}. ML enables the identification of complex, often non-trivial correlations between various turbine measurements, allowing for earlier and more precise anomaly detection. These ML models are typically classification models that, based on a set of measurements, classify the status of the turbine component as either “good” or “anomalous.” While such classifiers are already in use within the industry, our goal is to take it a step further and provide optimized strategies for (ii). Specifically, given a classifier and a status classified as ``anomalous," we aim to dynamically generate a control strategy that adjusts the turbine’s actuators (or more precisely, the set of statistical features) to restore it to a healthy state. This could result in an automated system that prescribes the minimum cost actions required to prevent failure and initiate timely, automatic preventive maintenance measures.

\subsection{Impact}

While our methods could be applied to any complex engineering system, we will here focus on transformers in offshore wind turbines. Wind turbine transformers act as a link between wind turbines and the transmission grid and their role is to step up the low output voltage from the generator to a higher voltage level for distribution.
The  transformer of the wind turbine has a controller---designed by the wind turbine manufacturer, a black-box for the turbine operator---that returns the system to a safe state in case of an emergency. For example, if the power production increases, the temperature at the transformer may exceed the safety threshold; therefore, the controller starts the cooling system to lower its temperature. 

Our primary contribution is to use Machine Learning and Operations Research techniques to design a new controller that:
\begin{itemize}
    \item can act promptly and prescribe a control strategy as soon as an anomaly is detected (and before the emergency alarm is activated);
    \item suggests an optimal control strategy that not only brings the component back to safety but does so minimizing changes;
    \item can be adapted to user preferences and application-specific constraints;
    \item is dynamic and tailored to a specific component and time instant.
\end{itemize}

Early anomaly detection using Machine Learning, combined with our optimized counterfactual controller, extends component lifetime and reduces reactive maintenance. Moreover, as discussed in Section \ref{sec:results}, determining the minimal parameter adjustments that are required to restore a healthy state is nontrivial. As we will see in Section \ref{sec:results}, traditional controllers are often overly conservative, whereas our approach identifies the optimal strategy based on real-time turbine measurements, effectively modeling their complex interdependencies.

A key contribution of our methodology is its ability to determine the best curtailment strategy, tailored to both the specific time and turbine type. Additionally, as shown in Section \ref{sec:accuracy&CE}, our approach allows for tuning based on the business case of individual farms and the user's risk perception. By quantifying the trade-offs between conservative and aggressive control strategies, our method facilitates a more informed discussion on farm-specific controller optimization.


Overall the main contributions of this paper are:
\begin{itemize}
    \item A novel application of Operations Research methods to counterfactual analysis: While most counterfactual analysis research focuses on individual decision-making scenarios (e.g., loan approvals, medical diagnoses), we introduce optimized counterfactuals in a new setting---fault prevention in complex energy systems.

    
    \item A novel mixed-integer quadratic programming model for counterfactual control, incorporating nested Machine Learning models.
    \item A practical contribution to fault detection and controller strategies, introducing a new methodology for optimal control based on counterfactual analysis, yielding potential increased revenues in the order of 3 million \euro~ per year in a typical farm.
    \item Real-world validation using real data and domain expertise from a leading wind energy company
    \item The use of Machine Learning functions, embedded within our mathematical optimization model, to estimate unknown processes (which are inaccessible, in our case, due to privacy reasons).
\end{itemize}

The rest of this paper is organized as follows: Section \ref{sec:ML} starts by describing the real-world data of the target problem and proposes a Machine Learning classifier for counterfactual analysis. Section \ref{sec:mathModel} defines our mathematical model for a generic optimal counterfactual control, while subsection \ref{subsec:controller simulation} enriches the mathematical model to ensure feasibility and safety for our specific application. Section \ref{sec:results} shows how our counterfactual controller acts on real data and analyzes the differences with the controller currently at the turbine. The next sections are dedicated to extensions to the problem. In particular, Section~\ref{sec:accuracy&CE} investigates the correlation between the accuracy of our ML model and the controller behavior, and how this can be used to fit user preferences. Section \ref{sec:cost-opt} researches the impact of considering cost in the objective function of our counterfactual optimizer. Finally, Section \ref{sec:conclusions} draws some conclusions and provides new directions for future work.

\section{Machine Learning models for classification of anomalies in the  turbine transformer} \label{sec:ML}

This section discusses the development of the proposed Machine Learning model, which is based on company data and expertise.

\subsection{Real-world data collection} \label{subsec:data}
We used data provided by Vattenfall BA Wind, a leading wind energy company in North Europe. We focus on the behavior of an oil-type transformer in a real-world farm. 
The data refer to signals from a real-life turbine measured every 10 minutes for about 4 years.

\begin{figure} [H]
     \centering
     \begin{subfigure}[b]{0.43\textwidth}
         \centering
         \includegraphics[width=\textwidth]{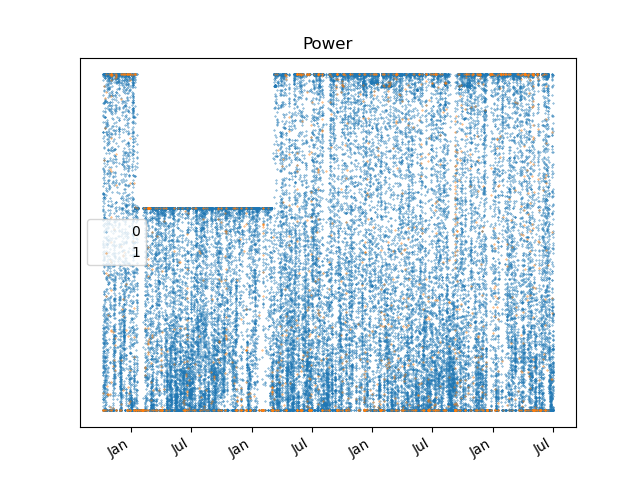}
         \caption{Generated power}
         \label{fig:power}
     \end{subfigure}
     \begin{subfigure}[b]{0.43\textwidth}
         \centering
         \includegraphics[width=\textwidth]{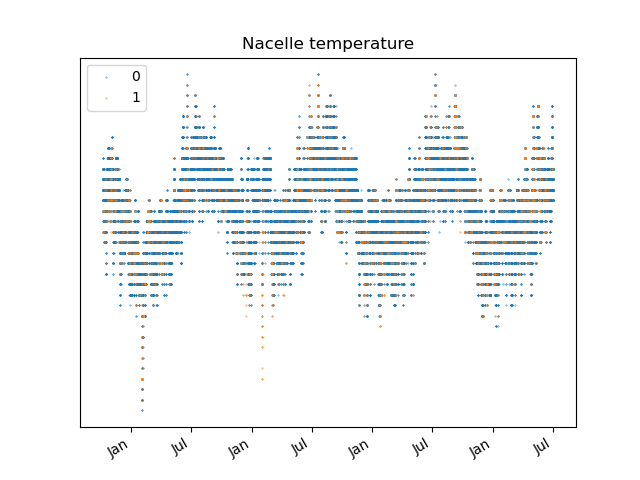}
         \caption{Temperature of the nacelle}
         \label{fig:temp_necelle}
     \end{subfigure}
     \hfill
     \begin{subfigure}[b]{0.43\textwidth}
         \centering
         \includegraphics[width=\textwidth]{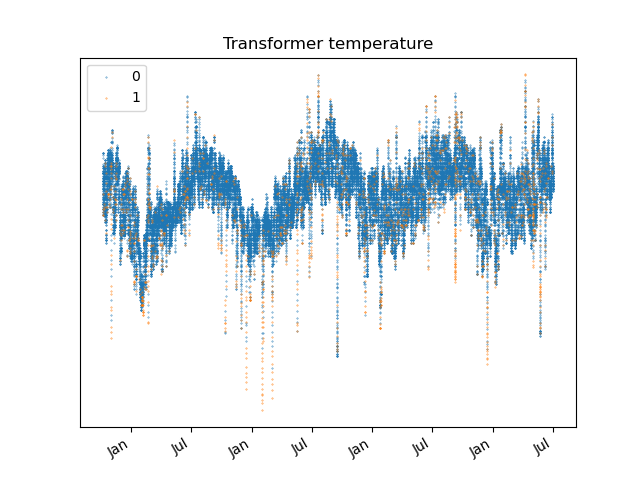}
         \caption{Temperature of the transformer}
         \label{fig:temp_transf}
     \end{subfigure}
     \begin{subfigure}[b]{0.43\textwidth}
         \centering
         \includegraphics[width=\textwidth]{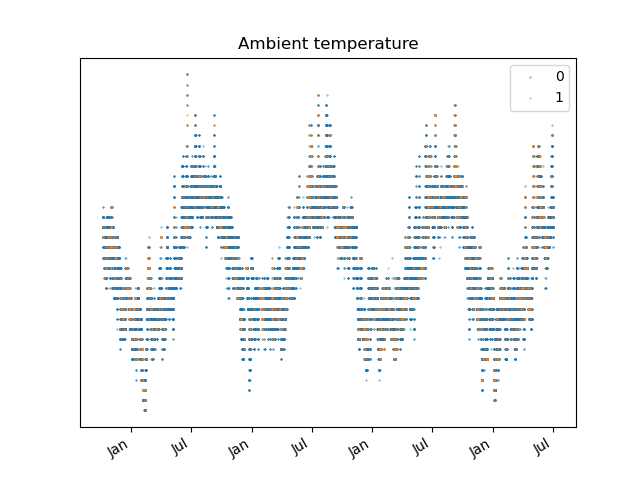}
         \caption{Ambient temperature}
         \label{fig:ambient}
    \end{subfigure}
     \begin{subfigure}[b]{0.43\textwidth}
         \centering
         \includegraphics[width=\textwidth]{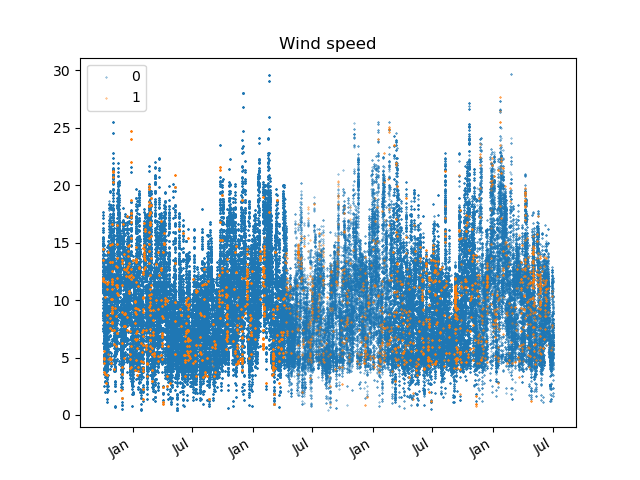}
         \caption{Wind speed}
         \label{fig:ambient}
    \end{subfigure}
        \caption{Input features over time. The data correspond to real measurements and come with a label that indicates whether the behavior of the component was normal (blue) or abnormal (orange). The y-axis values are hidden for privacy reasons. The data is collected over four years.}
        \label{fig:input_data1}
\end{figure}

The input data comprise different features and labels that indicate whether the measurement was defined as an \textit{anomaly} or not according to the company rules. \\
The considered input features are the following, all measured at each time-stamp:
\begin{itemize}
    \item The power production of the turbine. 
    \item The temperature of the transformer. 
    \item The nacelle temperature (i.e., the temperature measured in the room where the transformer is).
    \item The ambient temperature (i.e., the temperature outside the turbine).
    \item The wind speed. 
\end{itemize}
The input values are plotted in Figure \ref{fig:input_data1}. The y-axis is hidden for privacy reasons.

Due to the nature of the problem, our dataset is highly unbalanced, with the majority of samples labeled as ``good" (47\,883) and only a small fraction tagged as ``anomalous" (1\,759). Figure \ref{fig:input_data1} displays the feature distributions, where normal behavior (label 0) is shown in blue and anomalies (label 1) are shown in orange. In Section~\ref{sec:undersampling}, we discuss our approach to handling this imbalance during model training and testing.

These features are essential for our machine-learning classifier as they capture key aspects of the system: component state (transformer and nacelle temperature), turbine status (power production), and environmental conditions (ambient temperature and wind speed).

In addition, we found that incorporating a feature that captures the relationship between wind speed and power production improved the model’s performance. Therefore, we included the maximum power production at a given wind speed in our feature set.


Thus, for all input data, we computed the maximum allowable power production for the given wind speed, denoted as $\Pmax(\xws)$.
This value will be used in two key ways:
\begin{itemize}
    \item As an input feature for the Machine Learning model: Comparing $\Pmax(\xws)$ with actual power production helps identify anomalous behavior.
    \item As a constraint in our optimization model: Ensuring that the optimized controller never suggests exceeding the warranted power curve.
\end{itemize}

\subsection{Machine Leaning classifier}

As mentioned earlier, the use of Machine Learning models for fault detection is becoming a standard practice in the industry. Our optimized counterfactual control strategy can be applied to these models, providing companies with valuable new insights.
In this paper, we develop our own Machine Learning model, designed to replicate the one internally used by the company, and we train it using real-world data.
Indeed, we received a dataset from the company, where each sample is labeled as either ``good" or ``anomalous." These labels are internally defined by the company using Machine Learning techniques. Their approach involves training models on the transformer's normal behavior to predict its expected future performance. When actual measurements deviate significantly from these predictions over a sustained period, the instance is labeled as an anomaly. We use these labels to train our classifier. 

In this section, we discuss the Machine Learning model that we have used as a classifier in our mathematical model for optimal  control based on counterfactual analysis.

\subsubsection{Definition of the training and test sets} \label{sec:undersampling}

We split our data into training, validation, and testing sets, ensuring that we learn from the past and test on the future. To achieve this, we identify the timestamp at which 70\% of the anomalies had occurred. Data up to this point is used for training and validation (``past''), while data beyond this point is reserved for testing (``future'').
For our specific transformer, this cutoff occurs on August 20 of the 3rd year in our dataset—--data before this date is used for training, and data after it for testing.

\begin{figure} [!h]
     \centering
     \begin{subfigure}[b]{0.43\textwidth}
         \centering
         \includegraphics[width=\textwidth]{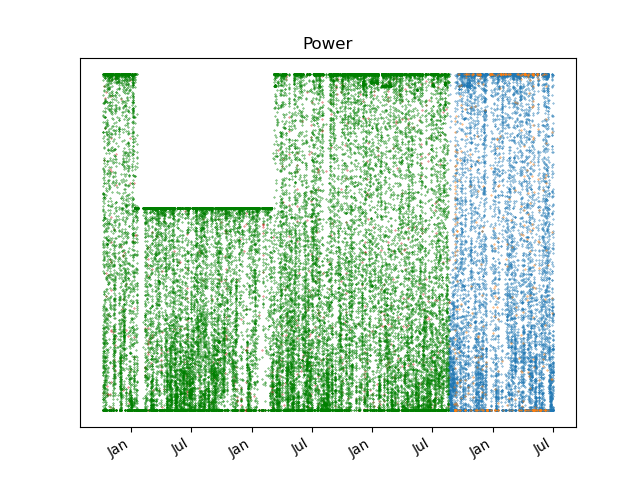}
         \caption{Generated power}
         \label{fig:power}
     \end{subfigure}
     \begin{subfigure}[b]{0.43\textwidth}
         \centering
         \includegraphics[width=\textwidth]{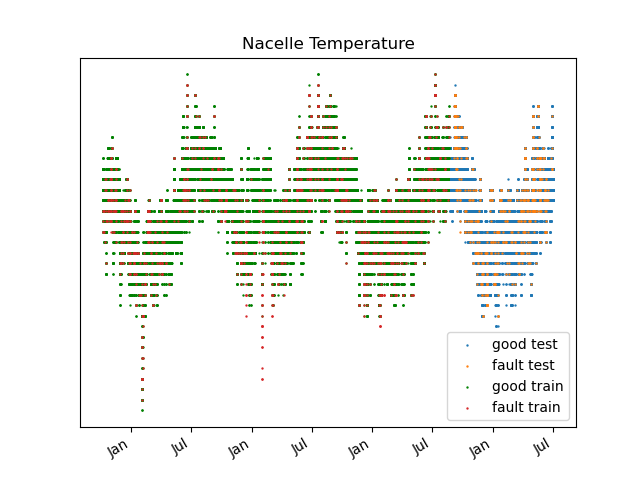}
         \caption{Temperature of the nacelle}
         \label{fig:temp_transformer_cell}
     \end{subfigure}
     \hfill
     \begin{subfigure}[b]{0.43\textwidth}
         \centering
         \includegraphics[width=\textwidth]{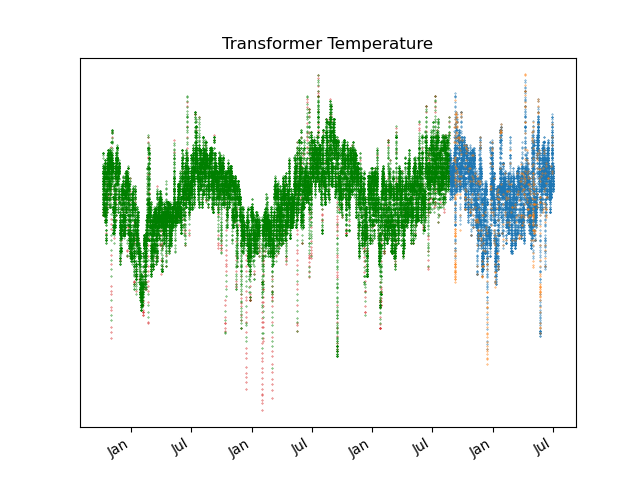}
         \caption{Temperature of the transformer}
         \label{fig:temp_transf}
     \end{subfigure}
     \begin{subfigure}[b]{0.43\textwidth}
         \centering
         \includegraphics[width=\textwidth]{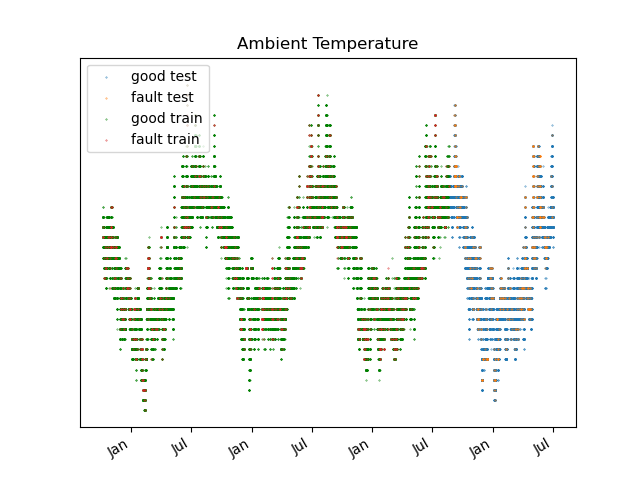}
         \caption{Ambient temperature}
         \label{fig:ambient}
     \end{subfigure}
        \caption{Transformer data used to build our Machine Learning models and their split into training/validation and test sets: in the training set, we use green to indicate the points labeled as good and red for the ones labeled as anomalies; in the test set, we use orange for the good points and blue for the anomalies. The values in the $y$-axis and the exact years are hidden for privacy reasons.}
        \label{fig:input_data}
\end{figure}

Figure \ref{fig:input_data} shows the split into training and testing for the different input features. Colors are used to show the labels for each data point: in the training set, we use green to indicate the points labeled as good and red for the ones labeled as anomalies; in the test set we use orange for the good points and blue for the anomalies.

Due to the nature of the problem, it is apparent that our dataset is highly unbalanced, with the "anomaly" category underrepresented compared to the "good" category. Training a Machine Learning model on such data risks producing a model that achieves high accuracy simply by always predicting ``good," without truly learning to distinguish anomalies.

To address this imbalance, we apply under-sampling when defining our training set. Specifically, we include all faulty measurements and randomly sample an equal number $n$ of ``good" measurements from the available data. This results in a balanced training set with an equal distribution of normal and faulty cases.

We leave the test set unbalanced to ensure it reflects real-world conditions as accurately as possible.

\begin{figure}[!h]
\centering
\includegraphics[width=0.5\linewidth]{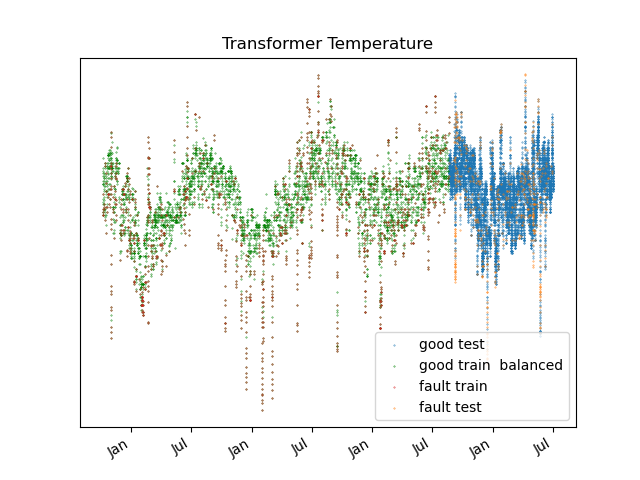}
\caption{\label{fig:undersample} Under-sampling technique to balance the training set: By comparing with \ref{fig:temp_transf} one can visualize the effect of under-sampling the good data in the training set (less green points, but the same number of red points). }
\end{figure}

In Figure~\ref{fig:undersample} we plot the so updated training and test sets for one feature (transformer temperature). By comparing this figure with sub-figure \ref{fig:temp_transf} one can notice the effect of under-sampling on the training dataset.

The balanced training set we have constructed in this way allows us to check if our ML model is indeed learning from the data.

\subsubsection{Machine Learning models} \label{subsec:NN}


We trained a Neural Network (NN) using ScikitLearn libraries to mimic the company Machine Learning classifier. 

The initial step of our controller involves assessing whether the feature sample is classified as anomalous.
Using the features from the transformer described in Section \ref{subsec:data}, the NN determines whether the component's status is normal or anomalous.  If  anomalous, the sample is then forwarded to our optimal counterfactual controller, which will establish a new status for the component that resembles its current status as close as possible, yet brings the system back to a normal state (see Section~\ref{sec:mathModel}). 

In our NN we used as features the signals from the wind turbine (i.e. power production, temperature of the transformer, temperature of the nacelle, ambient temperature, wind speed) as well as the additional feature $\Pmax$ for each sample.
Given the unbalanced nature of our problem, we use the balanced accuracy score to optimize the parameters of our NN. According to our experiments, the same confusion matrix results no matter if we optimize the $f_1$ score or the average precision score. The reader can refer to, for example, \cite{PerfMeasurments} for the definition of these scores. Furthermore, the hyper-parameters of our NN are estimated using $K$-fold cross validation.

\begin{figure}[!h]
\centering
\includegraphics[width=0.5\linewidth]{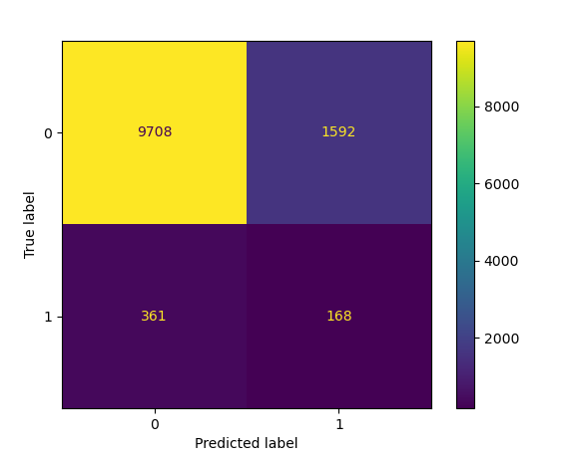}
\caption{\label{fig:CM} Confusion Matrix for the Neural Network classifier}
\end{figure}

The so obtained NN exhibits a standard accuracy of 66\%, an average precision score of 58\% and a balanced accuracy of 22\%.
The confusion matrix for our NN is particularly useful to understanding the behavior of our ML model, see Figure~\ref{fig:CM}.

Each time our application predicts a ``1'', i.e., an anomaly, our counterfactual optimization model is run, and a control strategy is set. Our experiment section will show that this approach does not incur a high expense for the company. In contrast, failing to identify issues, i.e., predicting ``0'' when the actual value was ``1'', can pose  a much greater risk. Through discussions with practitioners from the company, we have determined that, in our specific scenario, it is preferable to err on the side of caution. This means accepting more false positives (top-right quadrant of the confusion matrix) rather than false negatives (bottom-left quadrant), which could lead to overlooked issues. Given this, we are satisfied with our neural network’s performance. A company could take this even further by minimizing false negatives even more. We will explore the relationship between model precision and counterfactual control optimization in Section~\ref{sec:accuracy&CE}.


\section{Mathematical model for fault-control counterfactual optimization} \label{sec:mathModel}

The counterfactual-based optimization controller is run every time the Machine Learning classifier $f$ we have trained identifies an instance $x^*$  for which $f(x^*)$ = 1 (i.e., instance $x^*$ is predicted as anomalous).

The set of variables $x$ defines the status of the counterfactual control solution. The features that compose it are indicated as TN (temperature nacelle), TA (ambient temperature), TT (transformer temperature), P (power production), WS (wind speed), D (date, i.e. year, month, day, hour, minutes). To simplify notation we will indicate a generic feature as $x_j$, with $j \in [\text{TN, TA, TT, P, WS, D}]$. 
Note that the original instance that activates the controller have the same features, i.e. 
$ x^{*} = [ \xsp , \xstn, \xstt, \xsws, \xsd, \xsta ]$
 with $f(x^*) = 1$ (the current status is classified as anomalous).
 
We write a mathematical model to determine a $x = [ \xp , \xtn, \xtt, \xws, \xd, \xta ]$ that is the most \emph{similar} to $x^*$, but for which  $f(x)$ = 0 (that is, instance $x$ is predicted as good). We say that $x$ is the optimal counterfactual of $x^*$.

Importantly, due to the nature of our problem, we can alter the value of some variables while we cannot change others. Let $\mathcal{C} $ represent the set of controllable variables $\mathcal{C}  = [ P, TN, TT ]$, and with $\mathcal{F} $ the set of ``fixed'' variables $\mathcal{F}  = [TA, WS, D] $. The optimizer takes the following mathematical form:

\begin{align}
   &\min\sum_{j \in \mathcal{C} } \frac{1}{\overline{x}_j^2}  \left(x_j^* - x_j\right)^2  \label{eq:obj}\\ 
   &\text{subject to:} \nonumber \\ 
   & y = f(x) \label{eq:ML}  \\
   & y \leq s(\tau) - \epsilon \label{eq:classif}  \\
   & \lb_j\leq x_j \leq \ub_j\  ~~~~~~ \forall j \in \mathcal{C} \setminus\{P\} \label{eq:fesib} \\
   & x^*_j \leq x_j \leq x^*_j  ~~~~~~ \forall j \in \mathcal{F} \label{eq:fixed} \\
   &  x_j \in Z ~~~~~~ \forall j \in \mathcal{C}  \label{eq:integer}
\end{align} \label{model:counterf1}

where:
\begin{itemize}
    \item $s(\tau)$ is the score of the neural network without the sigmoid layer corresponding to the threshold probability $\tau$. For example, here $\tau =0.5$ corresponds to $s(\tau) = 0.0$.
     \item $\epsilon$ is a confidence value around the threshold. 
    \item $ \lb _j$ and $ \ub _j$ are the minimum and maximum allowable value for the $j$-th feature, respectively. These can be given by practitioners or computed based on historical data. 
    \item $ \avg _j$ is the average value of the $j$-th feature of $x$ computed over  all the dataset, and is used for rescaling the feature components.
\end{itemize}

The objective function \eqref{eq:obj} minimizes the squared Euclidean distance between the current status of the component $x^*$ and the counterfactual $x$. As the different features can have different magnitudes, we normalize their values by dividing them by $\avg$ (in order to give the same importance to all the features).
Constraint~\eqref{eq:ML} provides the predicted class $f(x)$, i.e., $y$, as per our ML model $f$; while constraint \eqref{eq:classif} ensures that the counterfactual is predicted to belong to the class 0 (i.e., not an anomaly). 
We used the Gurobi Machine Learning library \citep{gurobiML} in Python to code the model above, using Gurobi 12.0.1, applied to our scikit-learn~\citep{scikitLearn} neural network. For implementation purposes in Gurobi, $f$ is our NN classifier without the last layer, i.e., $y$ is the probability of being 0 or 1 (and thus, a number between 0 and 1), and $s(\tau)$ is the threshold value used in the last layer of the NN to determine whether the input feature is finally classified as 0 or 1. In constraint~\eqref{eq:classif} we added a confidence value $\epsilon$ to be sure that our counterfactual belongs to the ``good" class with some margin.
Constraints~\eqref{eq:fesib} set the feasible region for the features that we can change in our dataset. In our case, we define this region by looking at the minimum and maximum value that every feature in $C$ takes on in the whole dataset. One could also use domain-specific knowledge to define this region (as we will do for power in Section \ref{subsec:controller simulation}). Constraints \eqref{eq:fixed} fix the features that we cannot change (ambient temperature, wind speed, time stamp, etc.).
Finally, constraints~\eqref{model:counterf1} imposes that the counterfactual be integral. This is a requirement from the company related to the precision of the controller they would use to implement the counterfactual in practice.

The model above is of general application for fault control strategies. Our specific application to wind turbines required us to consider also some nontrivial dependencies between features, and some additional limitations, through additional constraints that we detail in Section \ref{subsec:controller simulation}

\subsection{Mathematical optimization for feasible and safe wind-turbine counterfactual control} \label{subsec:controller simulation}
The real-world application of our model \eqref{model:counterf1} requires not only optimizing the best preventive strategy to reduce failures but also ensuring the component’s safety at all stages. To achieve this, we need to incorporate application-specific constraints.

\subsubsection{Power curve}
The first constraint pertains to the warranted power curve.
Each turbine model, indeed, has a specific \emph{power curve}—--a function that relates wind speed to power production, $\Pmax (\xsws)$. This function is unique to each turbine, as it depends on factors such as rotor diameter and overall dimensions, and is typically provided by the manufacturer. Generally, power production increases with wind speed but eventually saturates beyond a certain point. The maximum power a turbine can generate is known as its \emph{rated power}.

\begin{figure}[h!]
    \centering
    \includegraphics[width=0.5\linewidth]{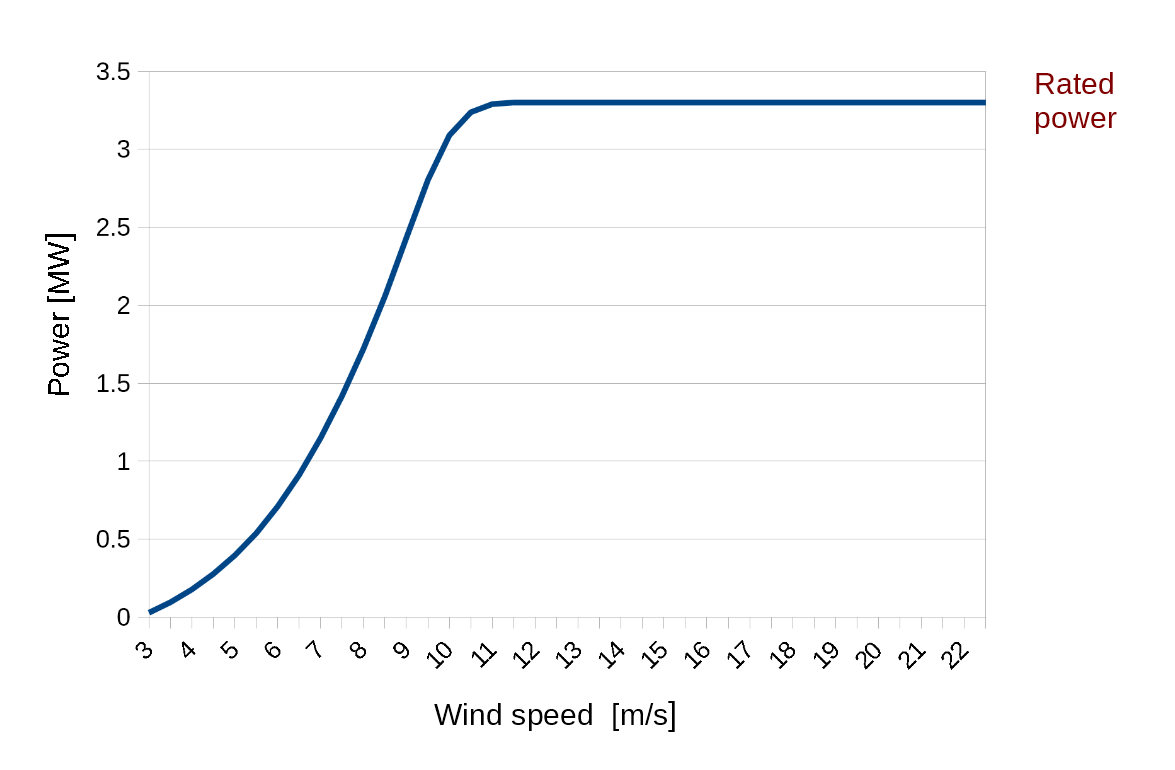}
    \caption{An example of a power curve showing the (non-linear) relation between power production and wind speed (source: \cite{FISCHETTI2019289}).}
    \label{fig:powerCurve}
\end{figure}

For privacy reasons, we cannot disclose the exact power curve used in our analysis. However, Figure~\ref{fig:powerCurve} presents a publicly available power curve of a small turbine--- from \cite{FISCHETTI2019289}---as an example. The characteristic S-shape of this curve is common across all wind-turbine power curves.

Operating a turbine above its designated power curve is risky and never done in practice. However, it is possible to operate below the power curve---a practice known as \emph{curtailment}. Since energy production is the primary business objective, turbines typically operate at maximum capacity. However, curtailment strategies can be employed when necessary to ensure turbine safety.

All in all, we need to include in our model that the turbine production can never exceed $\Pmax (\xsws)$, as defined by the manufacturer, i.e., 
   $$ \begin{aligned}
   & 0 \leq x_{P} \leq \Pmax( \xsws) \\
   \end{aligned} \label{model:MAX_P}
   $$
\subsubsection{Power -- temperatures relation}

Capturing the relationship between power production and transformer's temperature is a more complex challenge. This is governed by an internal black-box controller to protect the transformer’s mechanical integrity. 

When a turbine is sold, the manufacturer installs all controllers, but the buyer (i.e., the turbine operator) may have limited access to them or perceive them as black boxes. In our case, we have access to the power controller, which allows us to set the turbine’s power output. However, the temperature controller remains opaque to us. This controller, which relies on multiple sensors embedded in critical components, regulates cooling to prevent overheating —especially during high power production— thereby avoiding emergencies. While it indirectly links temperature and power production, we cannot control it directly.

In practice, if an anomaly is detected in the transformer, we can adjust power production, which can impact other system variables, specifically, transformer and nacelle temperatures, but cannot directly alter these temperatures.

Understanding the relationship between power production and other system parameters is crucial for our counterfactual model to ensure that our optimizer does not suggest dangerous or unfeasible states. Since the temperature controller is a black box with proprietary internal workings, we use Machine Learning to estimate and model these relationships.

This issue, stemming from a practical need at our industrial partner, proves to be also an interesting example of how Machine Learning plays an extra role in our optimization model. 

Specifically, we trained two Machine Learning models, $n$ and $t$, which represent the safe nacelle temperature and the safe transformer temperature, respectively. These models estimate the temperature limits imposed by the black-box controller from the manufacturer, based on the other system parameters.

In other words, we add the following constraints to our model in Section~\ref{sec:mathModel}:
   $$ \begin{aligned}
   & \xtn = n(\xta, \xp, \xtt,\xws, \xd) \\
    & \xtt = t(\xta, \xp, \xtn, \xws, \xd) \\
   \end{aligned} \label{model:countroller}
   $$

With this extended model, we make sure that our optimized control strategy is implementable by the current controller and safe for the component.

We used XGboost to estimate \textit{n} and \textit{t}, with root mean square error (RMSE) of 0.58544 and 1.14231 respectively. This proves that it is possible to reverse engineer the controller behavior and accurately add it to our model. Figures \ref{fig:controller} visualize the predicted values of \textit{n} and \textit{t}.

\begin{figure} [!h]
     \centering
     \begin{subfigure}[b]{0.43\textwidth}
         \centering
         \includegraphics[width=\textwidth]{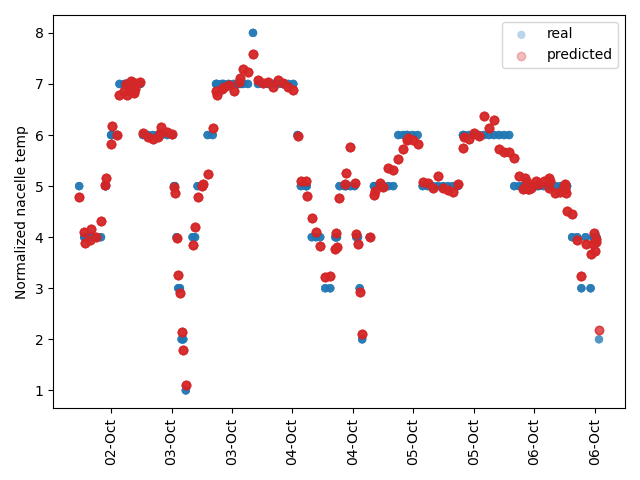}
         \caption{Nacelle temperature}
         \label{fig:n}
     \end{subfigure}
     \begin{subfigure}[b]{0.43\textwidth}
         \centering
         \includegraphics[width=\textwidth]{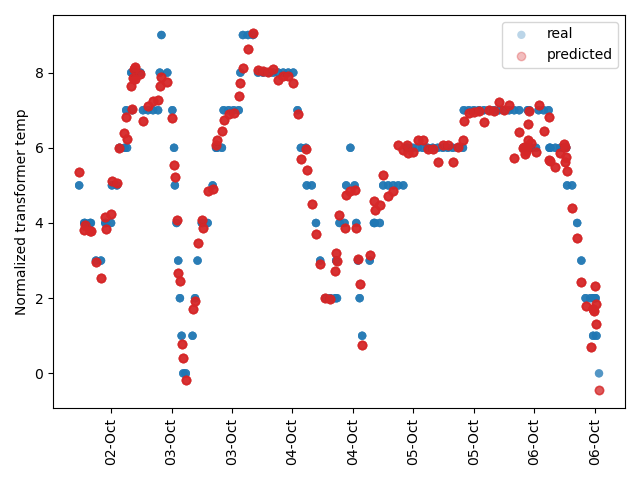}
         \caption{Transformer temperature}
         \label{fig:t}
     \end{subfigure}
        \caption{We use XGBoost estimators to mimic the black-box temperature controller behavior. The temperature values are normalized for privacy reasons}
        \label{fig:controller}
\end{figure}
\subsubsection{Final model}
Overall the mathematical model we implemented is as follows.
\begin{subequations}
  \begin{align}
   &\min\sum_{j \in \mathcal{C} } \frac{1}{\overline{x}_j^2}  \left(x_j^* - x_j\right)^2 \label{eq:obj_completeModel}\\ 
   &\text{subject to:} \nonumber \\ 
   & y = f(x) \\
   & y \leq s(\tau) - \epsilon \\
   & \xtn = n(\xta, \xp, \xtt,\xws, \xd) \label{eq:n}\\
    & \xtt = t(\xta, \xp, \xtn, \xws, \xd) \label{eq:t}\\
   & l_j\leq x_j \leq u_j\  ~~~~~~ \forall j \in \mathcal{C} \setminus\{P\} \\
  & 0 \leq x_{P} \leq \Pmax( \xsws) \label{eq:MAXP}\\
   & x_j = x^*_j  ~~~~~~ \forall j \in \mathcal{F}  \\
   &  x_j \in \mathbb{Z} ~~~~~~ \forall j \in \mathcal{C}  \label{eq:lastEq_completeModel}
\end{align} \label{model:completeModel}  
\end{subequations}

While the operator does not have access to the manufacturer's black-box controller, they can control the turbine's power production. This means that the operator can set $\xp$ according to the solution of \eqref{model:completeModel}, after which the manufacturer's black-box controller adjusts the temperatures accordingly. Although we cannot directly influence this relationship, it is crucial for us to capture it in our model, as it affects the classification of $x$.

One could also add an additional layer of complexity to our model by opening the black-box of the temperature controllers (equations \eqref{eq:n} and \eqref{eq:t}). If we were indeed using this model as the turbine manufacturer, we could let our optimizer define the optimal value for the temperatures within feasible regions. In Section~\ref{sec:results} we will run some tests trying to mimic the manufacturer's perspective.

\subsubsection{Drastic changes in temperature}

Finally, we examined the feasibility of drastic changes in the turbine's variables within the time frame available to the controller. According to company experts, there are no limitations on power changes (i.e., it is possible to reduce the turbine’s rated power to zero within 10 minutes). However, changes in transformer temperature must be constrained by a turbine-specific maximum temperature variation, which we will refer to as $\maxtchange$. This value is dependent on the cooling system of the oil-type transformer and varies based on the turbine model. As such, it should be treated as an input set by the user.

We will, therefore, incorporate the following constraints into model~\eqref{model:completeModel}:
   \begin{align}
   & \xtt - \xstt \leq \maxtchange \label{eq:MAX_TT1} \\
   & \xstt - \xtt \leq \maxtchange \label{eq:MAX_TT2} 
   \end{align}

Note that, according to our results, \eqref{eq:MAX_TT1}-\eqref{eq:MAX_TT2} are dominated by \eqref{eq:t}. Nevertheless, it is important to add this limitation when \eqref{eq:t} is relaxed, as for example in Section~\eqref{subsec:maunfact}.

\section{Results} \label{sec:results}
We use the NN of Section \ref{subsec:NN} in the mathematical model \eqref{eq:obj_completeModel}-- \eqref{eq:lastEq_completeModel} (i.e. including maximum power limitations and the simulators of the temperature controllers). The mathematical model was implemented in Gurobi 12.0.1 and run on a 11th Gen Intel(R) Core(TM) i7-1195G7 (16 GB RAM) laptop with a time-limit of 15 minutes for each instance. 

\subsection{Results on real-world data} 


\label{subsec:VF}
To show the impact of our methodology in practice, we used real data from our industrial partner. In particular, we simulated the use of our controller over a month, i.e. the full month of May of our 4th year of data, sampled every 10 minutes (for a total of 1197 instances). 
At every timestamp, the measurements from the turbine are passed to our ML model, which detects if there is an anomaly or not. If an anomaly is detected, our controller \eqref{model:completeModel} is activated and its optimized strategy is implemented. 

Out of 1197 instances, 758 are already predicted as good (requiring no controller activation), 32 are infeasible, and 407 are optimized by our controller. In practice, whenever model \eqref{model:completeModel} is infeasible, we revert to the company’s default strategy: turning off production. Note that, those few cases where model \eqref{model:completeModel} is infeasible arise because the company classifies turbine shutdowns as ``anomalous''—--since production is often cut to zero in response to safety concerns. Therefore, when our model is infeasible, it indicates that the only way to restore safe operation is by shutting down the turbine. These instances typically occur in low-wind conditions, where the controller has little room for optimization. 

\begin{figure} [!h]
     \centering
         \includegraphics[width=\textwidth]{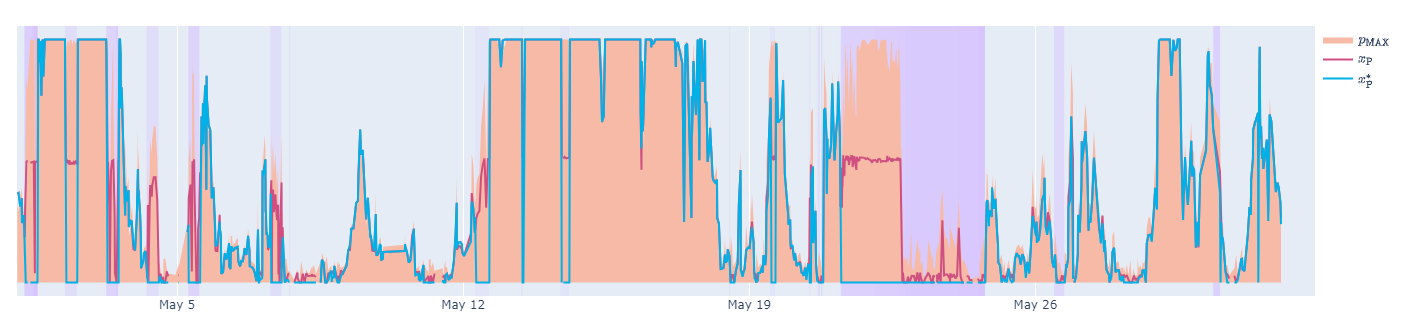}
         \caption{Power production: Comparison between the historical power production at the turbine (in blue) and the suggested power production by our counterfactual controller (in red). The pink area represents the feasible region according to the power curve and the current wind speed. The time intervals in which one or several alarms are active are highlighted in purple.}
        \label{fig:powerMayAlarms}
\end{figure}

As to the feasible instances, almost all the 407 instances are solved to optimality, and only 13 instances reach the time-limit (with a gap of 0.01\%).  
Figure \ref{fig:powerMayAlarms} shows the results: in blue the power production of the turbine according to historical data and in red the power production suggested by our controller. 
Note that the plot is actually an interpolation of a scatter plot in which, at every time-stamp, we register the status of the turbine (in blue) and, in case of an anomalous state, we compute the optimized new counterfactual state in red (that is as close as possible to the current blue state, but  is classified as healthy). When no anomaly is detected, the controller is not activated and only the blue line is visible. If both the blue and the red lines are visible, it means that our controller suggests a different strategy for the turbine.
As explained in Section \ref{subsec:data}, the production of the turbine is limited by the wind speed at that specific time and by the power curve of the specific turbine model: in Figure \ref{fig:powerMayAlarms} we plot the feasible area for power production in pink. It can be easily seen that the optimized production always lies in the feasible area.\\

Figure \ref{fig:powerMayAlarms} also shows  the main alarms detected by the safety system at the turbine in purple. Multiple alarms can be activated at the same time, the darker the purple, the more alarms are on.  

It is important to note that the current turbine controller is activated by an alarm system, and when active, it typically curtails power production to zero. This demonstrates that, once a problem is detected, it is not straightforward to determine the exact level to which power production should be reduced. Our data suggests that the current controller is quite conservative in its approach. For instance, on May 21st, after a problem was detected (indicated by the large purple area), the power production recorded by the turbine (in blue) drops to zero as a safety precaution. In contrast, our counterfactual controller shows that safety could have been maintained with a lower curtailment.

These differences in power production could lead to increased revenue for the wind energy company in the long run, while still ensuring the safety of the components. Assuming a revenue of 100 € per MWh (a realistic value proposed by our industrial partner, though not tied to any specific wind farm), we estimate that the company could have earned approximately 11\,500 € more in May by using our controller on the turbine under study. Thus in a typical park with 30 turbines, we can expect increased earnings in the order of 300\,000 \euro~ per month, i.e. more than 3 million \euro~ per year.

A fundamental aspect of our methodology is that our control approach is tailored to each instance and, via the integration of Machine Learning and Operations Research, is informed by all relevant features at the turbine at the precise point in time. Our approach does not involve a single, universally applicable strategy, but rather a dynamic methodology that identifies the most suitable control strategy based on the particular circumstances prevailing at the time of its application. This enables us to be less conservative and identify the most effective approach for the instantaneous measurements at the turbine, taking into account the intricate connections between the input features.

\begin{figure} [!h]
     \centering
         \includegraphics[width=\textwidth]{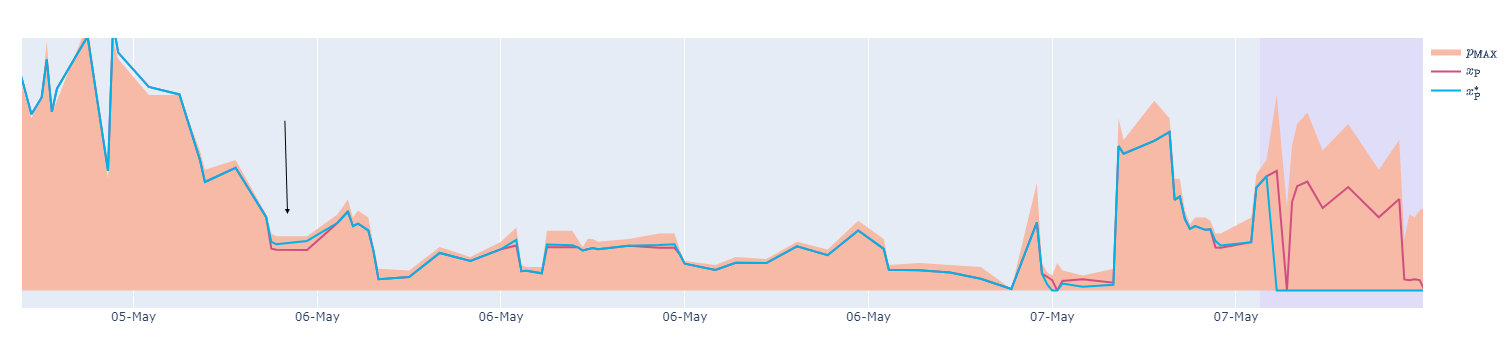} 
         \caption{ML can help detecting faults in advance and Operations Research can suggest the optimal reaction to move the component to a healthy state. Acting promptly could reduce damages at the transformer and increase the component lifetime.}
        \label{fig:powerMay57}
\end{figure}

Another key benefit of employing a controller that combines ML and OR techniques for fault detection lies in its ability to identify potential faults before they occur, unlike traditional alarm systems. Indeed, Machine Learning can help identify anomalies ahead of time, and Operations Research can provide the most effective strategy to bring the component back to a healthy state. 
An example of this can be seen in Figure \ref{fig:powerMay57}, where our Machine Learning model can identify anomalies earlier than the control system available at the moment at the turbine: Our counterfactual recommends curtailing already the 5th of May, while the controller currently at the turbine starts to act only on the 7th of May (when the alarm, in purple, is active). Acting promptly could reduce transformer damage and increase the lifetime of the component.

\subsection{The perspective of the turbine manufacturer} \label{subsec:maunfact}


As our work was developed in collaboration with a turbine operator, we focused on their perspective on the problem. The main effect of this is that our controller can only act on power production and, once the production changes, the black-box temperature controllers modify the temperature accordingly (constraint \eqref{eq:t}). Nevertheless, it is interesting to show that the same methodology can be applied by a turbine manufacturer, allowing the counterfactual to act directly on the transformer temperature as well. This is of practical interest (as it is motivated by the turbine-manufacturer perspective) and of scientific interest (as letting the optimizer act on more variables increases the complexity of our model).
Ideally, this should be done by replacing \eqref{eq:n} and \eqref{eq:t} with a mathematical model representing the feasible space for the temperatures. Unfortunately, this information is not shared by the manufacturers, not either with turbine operators. Therefore, in order to show the viability of our methodology for all the control systems, we estimated a possible feasible space by allowing a variation of +/- 10\% with respect to $\xtt$ from \eqref{eq:t} and to $\xtn$ from \eqref{eq:n}. In this way, we ensure that the variables are still related to each other, but we give our optimizer a bit of room to act on the temperatures as well.
To ensure feasibility with respect to sudden changes in temperature, we also enforce constraints \eqref{eq:MAX_TT1}--\eqref{eq:MAX_TT2}. In this test, we set $\maxtchange$ to 30 degrees.
We run the same test as in Section~\ref{subsec:VF}, with the same Neural Network model (of Section \ref{subsec:NN}).

\begin{figure} [!h]
     \centering
         \includegraphics[width=\textwidth]{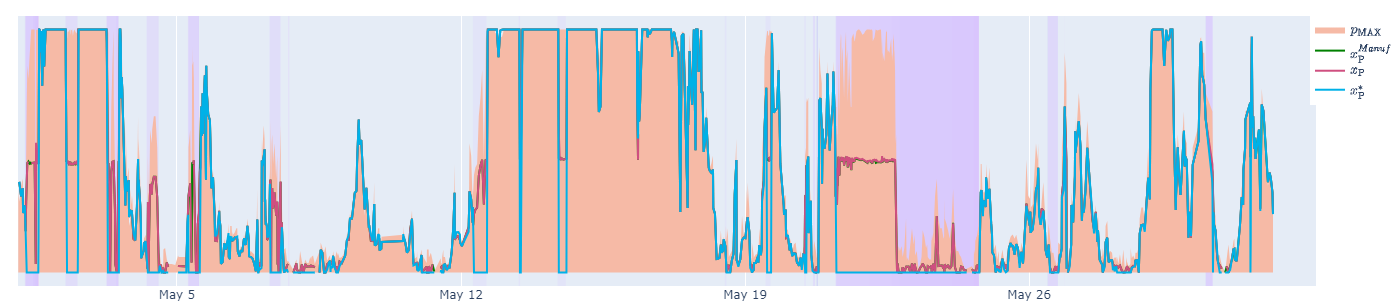}
         \caption{Power production: Comparison between the historical power production at the turbine (in blue) and the suggested power production by our counterfactual standard controller (in red) and the manufacturer-tuned one (in green). The pink area represents the feasible region according to the power curve and the current wind speed. The time intervals in which one or several alarms are active are highlighted in purple. The $y$-axis values are hidden for privacy reasons.}
        \label{fig:powerMayAlarmsM}
\end{figure}

Figure \ref{fig:powerMayAlarmsM} shows the results when allowing the optimizer to tune the transformer and nacelle temperature as well. 
Looking at Figure \ref{fig:powerMayAlarmsM} one can notice the impact on power production, which is minimal. 
Figure \ref{fig:controller} instead compares the transformer temperature. In red, the case from the turbine operator's perspective; in green, the case from the manufacturer's standpoint.
The two controllers mainly differ in the distance to the blue curve: the manufacturer controller (in green) changes less the temperature compared to the one in red. This is visible, for example, looking at the period (in purple) between May 19 and May 26. This is coherent with our model as, in the manufacturer's perspective, we allow more freedom to the transformer temperature that thus can minimize more the changes to the current status, according to our objective function.

\begin{figure} [!h]
     \centering
         \includegraphics[width=\textwidth]{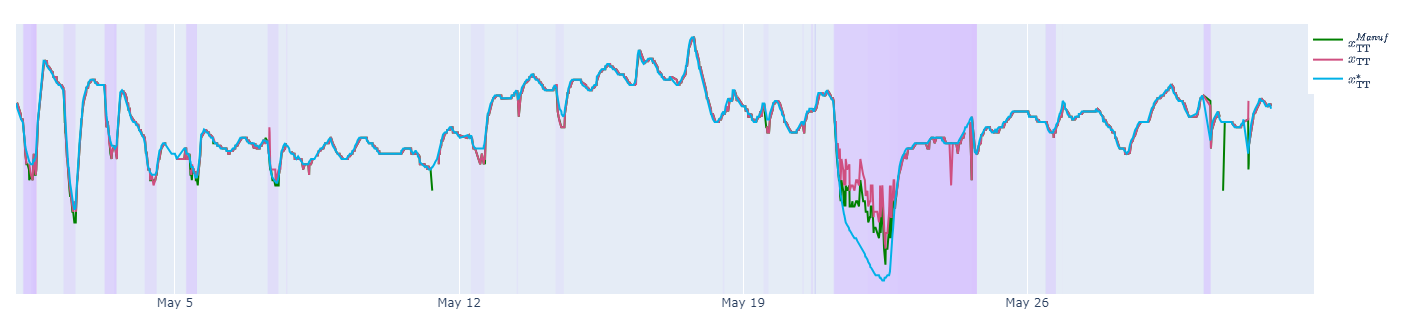}
         \caption{Transformer temperature: Comparison between the historical transformer temperature at the turbine (in blue) and the suggested one by our counterfactual standard controller (in red) and the manufacturer-tuned one (in green). The time intervals in which one or several alarms are active are highlighted in purple. The $y$-axis values are hidden for privacy reasons.}
        \label{fig:controller}
\end{figure}

\section{Considerations over ML accuracy and optimized Counterfactual controller: a way to consider user preferences} \label{sec:accuracy&CE}


An interesting aspect of our methodology is the relationship between the accuracy of the ML model and the control strategy, which can be leveraged to account for customer preferences. We find this particularly relevant, since, to the best of our knowledge, this connection remains unexplored in the literature, where counterfactuals have not been used for controllers.

As demonstrated in Section \ref{sec:results}, our OR+ML controller tends to adopt less conservative strategies compared to the one currently implemented in the turbine. Given that activating our controller does not lead to significant power losses, customers may be more inclined to use it frequently. In practical terms, a customer might prioritize detecting a higher number of anomalies rather than fewer, as early intervention incurs minimal costs with our controller and, conversely, failing to detect a fault could necessitate a turbine visit and costly maintenance operations. 
In terms of Machine Learning, this translates into accepting more false positives to reduce false negatives. This user-specific preference can be incorporated into the tuning of our ML model. Specifically, during the design of our under-sampling technique, we created a slightly unbalanced dataset, consisting of 25\% more faulty cases than good ones (i.e., overall the partition of our test set is 43\% good instances and 57\% faulty ones). Training the model with a higher proportion of anomalies compared to normal cases results in an ML model that places greater emphasis on detecting anomalies. All other training settings remain identical to the Neural Network described in Section \ref{subsec:NN}.

\begin{figure}[!h]
\centering
\includegraphics[width=0.5\linewidth]{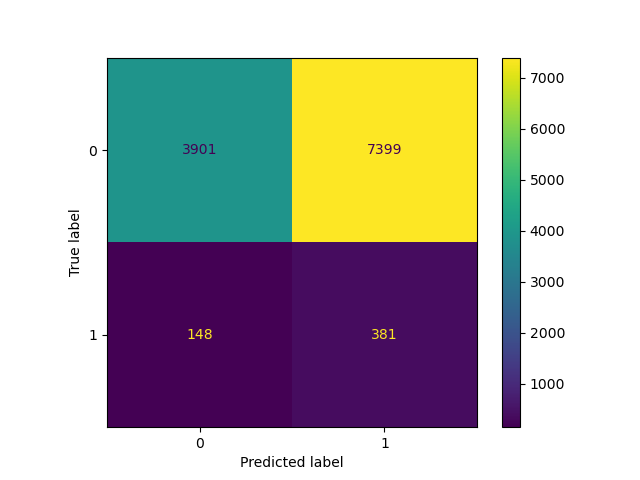}
\caption{\label{fig:CM025} Confusion Matrix for our Neural Network classifier that focuses on reducing false-negatives (bottom-left corner of the confusion matrix).}
\end{figure}

As a result, we obtained a Neural Network with a standard accuracy of 58\%, an average precision score of 61\%, and a balanced accuracy of 15\%. Most importantly, as evident from comparing the new confusion matrix (Figure \ref{fig:CM025}) with the previous one (Figure \ref{fig:CM}), the updated model exhibits a lower number of false negatives (bottom-left corner of the confusion matrix).

\begin{figure} [!h]
     \centering
         \includegraphics[width=\textwidth]{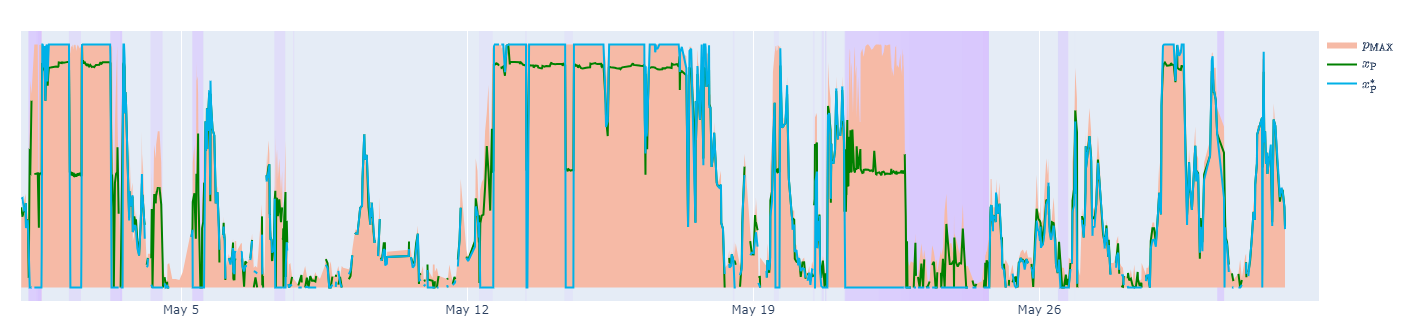}
         \caption{Power production: comparison between the historical of the power production at the turbine (in blue) and the suggested power production by our new counterfactual controller (in green). The pink area represents the feasible region according to the power curve and the current wind speed. In purple when alarms were active. The y-axis values are hidden for privacy reasons.}
        \label{fig:powerMayAlarmsM025}
\end{figure}

\begin{figure} [!h]
     \centering
         \includegraphics[width=\textwidth]{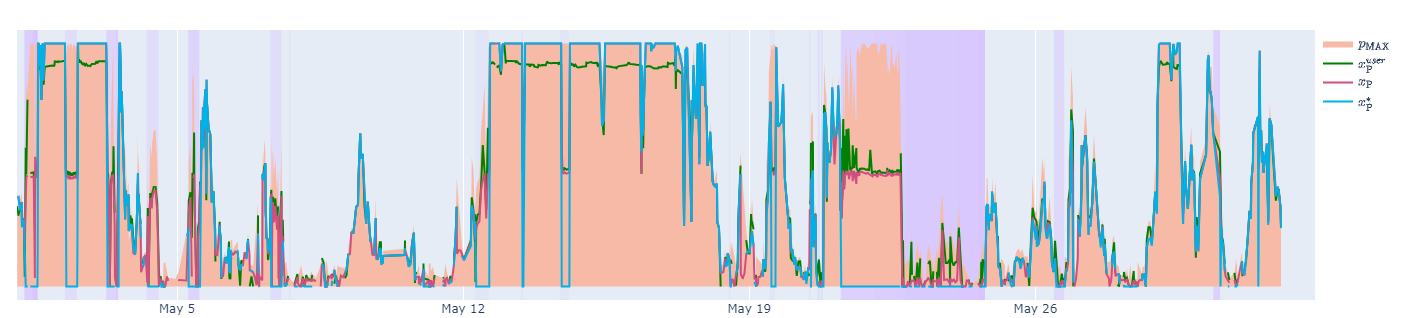}
         \caption{Power production: comparison between the historical power production at the turbine (in blue) and the suggested power production by our new counterfactual controller (in green) and the one by our controller with default settings (in red)}
        \label{fig:powerMayAlarmsM025C}
\end{figure}

We repeat the same tests from Section \ref{subsec:VF} using this new NN, trained according to user preferences (Figure \ref{fig:powerMayAlarmsM025}).
In Figure \ref{fig:powerMayAlarmsM025C} we can appreciate the impact of this different tuning of our NN. In particular, as expected, we can see that the counterfactual is activated more often (thus the green line more often differs from the blue one). 

If we assume a revenue for the company of 100\euro /MWh, we can estimate that, even using these more conservative settings for our neural network, our counterfactual control strategy still allows an extra gain of about 10\,500 \euro\  for the single turbine at hand.

Enabling companies in the wind energy sector to personalize the controller behavior in line with user preferences, and to measure resulting profit variations, is a highly effective asset. A tailored controller for each wind farm could be developed, taking into account the farm's unique business needs and risk assessment, rather than relying on generic strategies that are widely used in the industry (as is currently the case). 

\section{Revenue-driven counterfactual optimization} \label{sec:cost-opt}

Our initial counterfactual optimization aimed to minimize the change in component status required to achieve a healthy state.  However, in practice, other objectives may be important.  Maximizing revenue, for instance, is a compelling goal.
To be more specific, ideally one should maximize company income, balancing between immediate revenues and potential costs for failures.
Due to the difficulty of accurately quantifying potential failure costs, we implemented a two-stage optimization strategy intead.  First, we determined a baseline counterfactual, $\xr$, using model \eqref{model:completeModel}.  Second we maximized revenue while limiting the change in component status to a maximum of $\pi$ (10\% in our tests) from $\xr$.  This approach allows us to explore revenue maximization while mitigating the associated risk.

To do so, we change the objective function of model \eqref{model:completeModel} to 
$$ \max r \xp$$
(where $r$ is the energy price) and we add the constraint $$ \xp \leq \xrp + \pi  \xrp $$ 
In the following analysis, we assume a fixed energy price (as in the previous tests), as this is common for subsidized wind farms. However, our methodology is flexible and can also be applied to scenarios with variable or even negative energy prices. Indeed, as the model is dynamically re-run every 10 minutes on the new sensor measurements, one can easily update the value of $r$ to its (dynamic) value at each run. Note that in case of negative prices, the optimizer will suggest to lower the power production to the minimum.

\begin{figure} [!h]
     \centering
         \includegraphics[width=\textwidth]{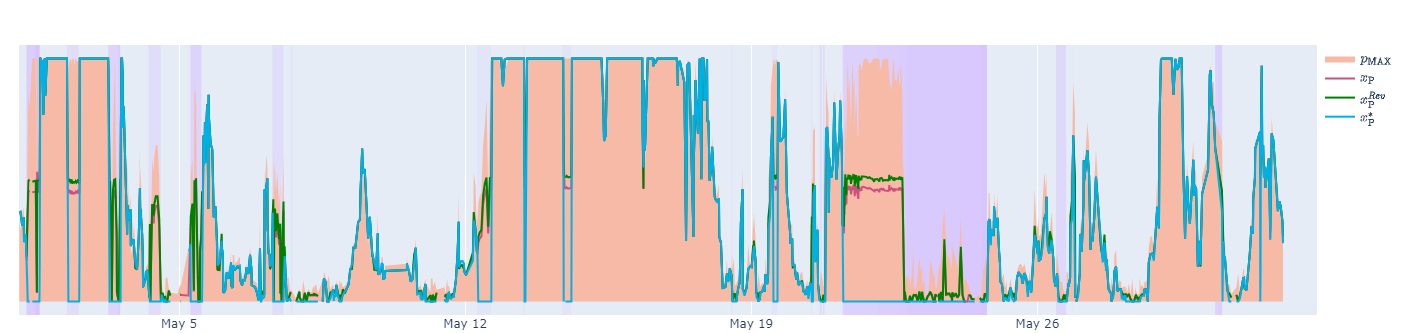}
         \caption{Power production: comparison between the historical power production at the turbine (in blue), standard counterfactual controller strategy (in red) and revenue-driven counterfactual controller strategy (in green). The pink area represents the feasible region according to the power curve and the current wind speed. In purple when alarms were active. The y-axis values are hidden for privacy reasons}
        \label{fig:POPTallMay}
\end{figure}

We run the new revenue-driven model on the same test-set of Section~\ref{sec:results}, obtaining the results highlighted in green in Figure~\ref{fig:POPTallMay} (indicated as $x_{\textrm{P}}^{Rev}$ in the caption). 

\begin{figure} [!h]
     \centering
        \includegraphics[width=\textwidth]{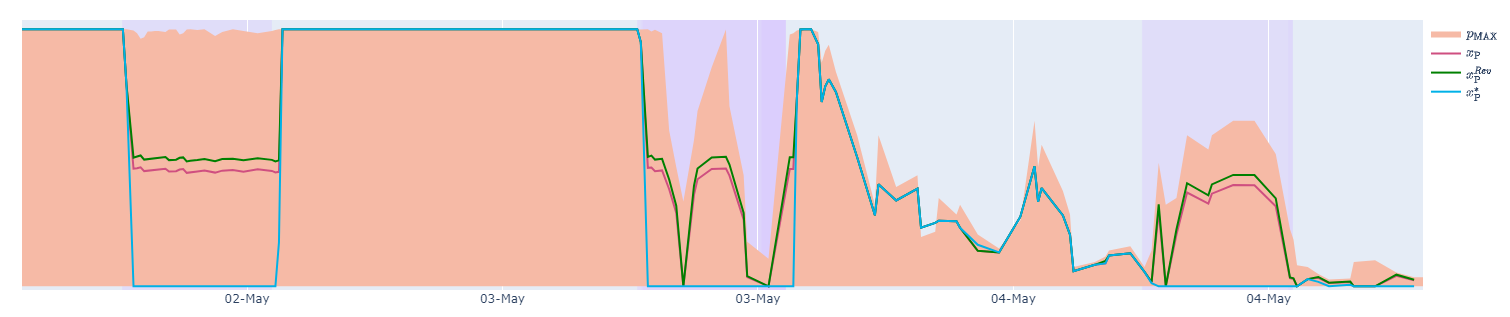}
         \includegraphics[width=\textwidth] {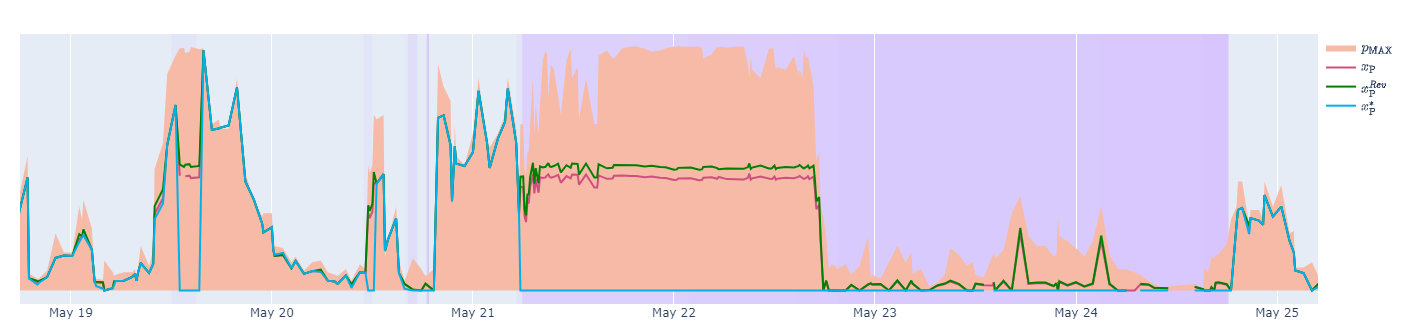} 
         \caption{Comparison of power productions with the different types of controllers. Zoom in the period 2-3 of May and 19-25 of May of the 4th year in our datatset}
        \label{fig:POPTallMayZOOM}
\end{figure}

We provide in Figure \ref{fig:POPTallMayZOOM} some zooms of the above plot to better appreciate the difference between the solution of the revenue-driven optimization (in green) compared to the standard optimization model (in red). If the red line is not visible, it means it is identical to the green one. Compared with the counterfactual control strategy suggested by model \eqref{model:completeModel}, in red, we can notice a further increase in power production. 

If we compute the extra revenue that the company could have obtained utilizing our revenue-driven controller strategy in the month of May (assuming the usual revenue of 100 \euro /MWh), this strategy is 12\,900 \euro~ more profitable than the one currently implemented. Looking instead at the extra gain compared with our counterfactual optimization from  \eqref{model:completeModel}, the extra gain in the month of May is of about 1500 \euro~ for the single turbine under study.

\section{Conclusions and future work} \label{sec:conclusions}

In this paper we have shown how Operations Research can be used in counterfactual analysis to prevent failures in complex energy systems. In particular, we have studied the oil-type transformer of a real-world wind farm, using real-world data, and we have designed an innovative type of controller that is both effective and flexible to user preferences. By integrating Machine Learning and Operations Research, our approach dynamically adapts to the turbine's specific conditions at any given moment: Rather than proposing a one-size-fits-all solution, we present a flexible method that continuously identifies the most effective control strategy based on real-time conditions and their intricate relationships.

Our results highlight the complexity of selecting the optimal control strategy to restore components to a healthy state. We show that mathematical modeling can yield more cost-effective solutions than traditional conservative control strategies, offering significant savings and flexibility to the user.

Our experiments demonstrate that our approach can be easily modified to accommodate user preferences. Specifically, we show that the method can be adapted based on the specific features that the user has the ability to act on (which may vary between a wind energy producer and a turbine manufacturer), it can be fine-tuned according to the user's level of risk-aversion, and the objective function can be altered to reflect business priorities (for instance, a revenue-driven controller). 
From the application side, our tests suggest that a company using our methodology could increase its production (i.e. revenue) significantly: looking at only one turbine during one month, we estimate savings in the order of 10\,000 \euro. Considering that offshore wind farms usually have about 30 turbines (old farms around 100 smaller turbines and newest farms around 20 bigger ones), this equals to savings in the order of 300\,000 \euro~ per month, i.e., more than 3 million \euro~ per year---these savings are naturally contingent on the number of anomalies/failures occurring during each month.

From a mathematical perspective, we show a new application of OR for counterfactual analysis, on a completely new setting. Driven by the application, we investigated how counterfactual optimization can deal with different objective functions and how nested Machine Learning models can be used in mixed integer quadratic programming models to estimate unknown functions. Interestingly, we also show how the accuracy of the Machine Learning model can have an impact on the optimization results, and how this can be used to incorporate user preferences. 

Our approach leverages OR and ML to provide timely solutions that minimize anomalies and faults. Notably, the same methodology can be adapted to address other objectives by playing with the label definition. For instance, a company might prioritize reducing faults that require a human visit to the turbine, as these interventions tend to be the most costly and hazardous. To consider this, the company would need to record the visits and simply connect each transformer-status sample to a label, equal to 1 if this fault required a visit, or 0 otherwise. By training the ML model on this new label, our existing mathematical models can be repurposed to develop a control system that minimizes turbine visits. More broadly, our methodology is highly adaptable---any aspect that can be labeled from the data can be incorporated, allowing the approach to be tailored to various operational priorities.

Another future work could be to include robustness considerations to the model. Indeed, one could consider counterfactual robustness with respect to the uncertainty of the ML model, and place more attention to the confidence value $\epsilon$. One could also consider a robust approach with respect to the input measurements (for example, saying that one of the sensor is less trustworthy than the other, or even considering an uncertainty buffer on all measures). Our methodology can be made robust with respect to these considerations, for example, by applying the methodology in \cite{Marango_Kurtz}. \\

\bibliographystyle{elsarticle-harv} 
\bibliography{my-biblio}

\section*{Acknowledgements}
This research has been funded under Grant QUALIFICA (PROGRAMA: AYUDAS A ACCIONES COMPLEMENTARIAS DE I+D+i) by Junta de Andalucía grant number QUAL21 005 USE and under the Ramon y Cajal grant number RYC2023-042623-I.

The work of J. M. Morales was supported by the Spanish Ministry of Science and Innovation (AEI/10.13039/501100011033) through project PID2023-148291NB-I00. 
The work of E.Carrizosa was supported by the Spanish Ministry of Science and Innovation through project PID2022-137818OB-I00.

From Vattenfall we would like to thank also Robbert van Kortenhof, Subhra Samanta and Adel Haghani, and Torsten Jeinsch from University of Rostock who developed the internal Machine Learning model for fault identification in Vattenfall (model on which we based our classifier) and Pawel Orlowski who helped us in our understanding of the controller dynamics.

\end{document}